\documentclass[aps,pra,epsfig]{revtex4}
\usepackage[dvips]{graphicx}
\usepackage{caption}
\usepackage{color}
\usepackage{epsfig}
\usepackage{soul}

\begin{document}

\title{Placing Kirkman's Schoolgirls and Quantum Spin Pairs on the Fano Plane: A Rainbow of Four Primary Colors, A Harmony of Fifteen Tones}
\date{\today}

\author{J. P. Marceaux and A. R. P. Rau$^1$}
\affiliation{$^1$Department of Physics and Astronomy, Louisiana State University, Baton Rouge, Louisiana 70803, USA}

\begin{abstract}
 
 A recreational problem from nearly two centuries ago has featured prominently in recent times in the mathematics of designs, codes, and signal processing. The number 15 that is central to the problem coincidentally features in areas of physics, especially in today's field of quantum information, as the number of basic operators of two quantum spins (``qubits"). This affords a 1:1 correspondence that we exploit to use the well-known Pauli spin or Lie-Clifford algebra of those fifteen operators to provide specific constructions as posed in the recreational problem. An algorithm is set up that, working with four basic objects, generates alternative solutions or designs. The choice of four base colors or four basic chords can thus lead to color diagrams or acoustic patterns that correspond to realizations of each design. The Fano Plane of finite projective geometry involving seven points and lines and the tetrahedral three-dimensional simplex of 15 points are key objects that feature in this study.   

\end{abstract}

%\pacs{03.65.Ta,03.67.-a}

\maketitle

\section{Introduction}

A problem presented as a simple puzzle about 175 years ago has had many interesting connections to several areas of mathematics and was recently linked to topics of current active ferment in the field of physics called quantum information. Posed in 1844 by Woolhouse  in an unusual but charming journal reference \cite{ref1}, the problem is to arrange 15 schoolgirls in five rows of three abreast for seven days of the week with no pairs in common in any row. Thus, no girl sees a repeat alongside in her row of three as the days pass by. The problem was solved by an amateur, Rev. Thomas Kirkman, who was himself a fine mathematician and gave a very general treatment going well beyond the recreational puzzle itself \cite{ref2,ref3}. Mainline mathematicians studied this problem of combinatorics as ``triple systems" \cite{ref4} and more general ``designs," especially what are called ``balanced incomplete block (BIB)" designs. A whole area of mathematics called Design Theory \cite{ref5}, important today in Coding Theory \cite{ref6}, traces back to these developments, and the use of BIBs in the design of experimental statistics goes back to Yates \cite{ref7}.

The great mathematical statistician R. A. Fisher formulated the basic principles of this subject while applying them to agricultural statistics and, together with finite geometers in an Indian Statistical Institute \cite{ref8}, established it as a rigorous mathematical topic \cite{ref9,ref10,ref11}. There is now a vast mathematical literature (for example, \cite{ref12,ref13,ref14,ref15}) arising from the ``Kirkman Schoolgirl Problem." One of them \cite{ref12} puts it in the following terms, ``The schoolgirl problem connects block designs, finite projective geometry, algebraic number fields, error-correcting codes, and recreational mathematics. Are there any other?" and proceeds to make one more, a connection to the Galois group of degree-16 field. In this paper, we make yet another, to a topic in quantum physics of a pair of quantum bits (``qubits") whose correlations form the basis for many envisaged applications in the quantum computers and cryptographies of the future. We use this connection and an algebra of triples satisfied by qubit operators to formulate Kirkman designs and realize them through patterns in color and sound.  

The number fifteen chosen by Woolhouse and Kirkman happens to be the number of generators of the group SU(4) and the number of operators of the su(4) algebra of two spin-1/2 systems in quantum physics and quantum information where logic gates, entanglement, and other quantum correlations require a pair of such spins as the minimal number for their description \cite{ref16}. With the recognition that the same aspect of triplets enters into either the multiplication (``Clifford algebra" \cite{ref17}) or commutation bracket (``Lie algebra" \cite{ref18}) of any two operators so as to form a closed algebra or group (along with zero and the identity operator), an intimate correspondence has been established between this Lie-Clifford algebra of quantum spin pairs and the Kirkman Schoolgirls \cite{ref19}; see also \cite{ref20}. We now develop this correspondence further to explore solutions of the Kirkman design in an abstract space of ``geometric algebra" \cite{ref21}, which generalizes the algebra of Lie and Clifford to geometric manifolds. We design a code that uses the algebraic properties of qubit operators to construct geometric mappings that fulfill the criteria of the Kirkman design.  

Fifteen is also the ``colors of the rainbow" in a four-color scheme just as seven colors of the rainbow, a familiar ditty from childhood, follow for our usual color vision based on three primary colors, red, blue, and green (or other variant names). Again, this is a simple consequence of the combinatorics of taking the basic objects and their combinations in pairs and higher multiples in possible different ways. As an interesting biological aside \cite{ref22}, while our human color vision is based on three, some birds, amphibians, and reptiles have four basic colors (cones in their retinas). This is a story of evolution. Mammals, who evolved from early ancestors with nocturnal habits, lost color vision that does not work in the dark (only the rods that sense light remaining) so that most mammals have only two-color vision whereas the earlier reptilian ancestors had four, retained in current birds and amphibians. The evolution of monkeys and later our primate ancestors regained a third color among some mammals, perhaps to recognize ripe fruits for their diet. Though human vision is really based on only three colors, the extraordinary range of hues that it can perceive allows us to find a set of fifteen colors that are visibly distinct so that they may form a representation for the schoolgirls in Kirkman's design. 

Our fifteen schoolgirls also make an appearance in music. While most western music is based on the canonical twelve-tone chromatic scale, musical scales can be constructed on any number of notes through different octave divisions. The composer Tom Johnson used such an alternative octave division in his musical composition ``Kirkman's Ladies" \cite{ref23}, a sonic representation of a Kirkman design system in which the schoolgirls or tones are arranged into sets of three as chords over a 15-note scale. Johnson's musical Kirkman design spans 7 intervals or ``days'' of 5 musical chords of three notes  or ``sets of three girls'' such that fifteen notes that form the chords obey the criteria of the Kirkman design system. At a notional level, musical composition strongly resembles the method of design construction. A composer arranges the distinct notes of the scale into blocks that form chords and then harmonies. 

The appearance of design in color and musical composition is the basis of our linking the mathematics of quantum spin pairs to Kirkman arrangements of sound and color bases. Interestingly, Newton's ``unweaving of the rainbow" into seven colors seems to have been based not on combinatorics of three primary colors or the much later recognition in biology of three retinal cone receptors but by analogy with musical sub-divisions of the octave. The structure of this paper is as follows. Section II gives a brief summary of the basic operators of two qubits, and of their Lie and Clifford algebraic structure on which we construct Kirkman arrangements and associated codes in Section III. This involves the Fano Plane of finite projective geometries \cite{ref5} together with an associated cube or, alternatively, a tetrahedral arrangement of the fifteen objects \cite{ref13}. An algorithmic construct allows various alternative solutions that satisfy the basic requirements posed by Woolhouse and Kirkman. Section IV maps the problem to a basic four-color scheme (really a 15-hue scheme) and Section V similarly to a fifteen-note scale. The algorithm can take a minimal set of qubit operators to output Fano and Kirkman arrangements in color or sound and we present examples.

\section{Pair of qubits, and their symmetries}

Any quantum system, even of a single particle or degree of freedom, exhibits the important property of ``superposition" that characterizes any linear theory. While already interesting in greatly expanding the states of a single qubit (all superpositions of $|0\rangle$ and $|1\rangle$, with complex coefficients of superposition), as compared to just two, 0 and 1, for a classical bit, it takes a minimum of two qubits to discuss all the topics of quantum information such as quantum logic gates, cryptography or secure key distribution, entanglement and other quantum correlations and, of course, the future quantum computer which will have a large number of qubits \cite{ref16}. A single qubit is described by the well-known Pauli matrices and their su(2) algebra or SU(2) unitary group symmetry and involves three basic operators or parameters. A pair of qubits, or two-level systems, requires a much larger number, fifteen, of generators or parameters, and is described by su(4) or SU(4). Various renderings of the fifteen and various sets of $ 4 \times 4$ linearly independent matrices can be used but one that is most convenient starts from the individual Pauli matrices, 6 in number, of the two qubits along with 
tensor products, 9 in number, rendered together as: $\frac{1}{2} \vec {\sigma} \otimes \mathcal{I}, \mathcal{I} \otimes \frac{1}{2} \vec 
{\tau}, \frac{1}{2} \vec{\sigma} \otimes \frac{1}{2} \vec{\tau}$. 
Together with the unit $4 \times 4$ matrix that we name $O_1$, these 16 matrices called 
$O_i$, with $O_2 =\frac{1}{2}\sigma_z, O_3=\frac{1}{2}\tau_z , O_4=\frac{1}{4}\sigma_z\tau_z, O_5=\frac{1}{2}\sigma_x, O_6=\frac{1}{2}\sigma_y, O_7=\frac{1}{4}\sigma_x\tau_z, O_8=\frac{1}{4}\sigma_y\tau_z, O_9=\frac{1}{2}\tau_x, O_{10}=\frac{1}{2}\tau_y, O_{11}=\frac{1}{4}\sigma_z\tau_x, O_{12}=\frac{1}{4}\sigma_z\tau_y, O_{13}=\frac{1}{4}\sigma_x\tau_x, O_{14}=\frac{1}{4}\sigma_y\tau_y, O_{15}=\frac{1}{4}\sigma_x\tau_y, O_{16}=\frac{1}{4}\sigma_y\tau_x$,  were explicitly tabulated in earlier papers \cite{ref24}.

When we discuss the construction of code in this paper, we also consider an alternative basis set written as $Q_n$ that is a simple linear map from the $O_i$ basis with a 1:1 correspondence. The $Q_n$ representation assigns a two-label binary string for each of the qubit operators: $\mathcal{I} : 00, \sigma_z : 01, \sigma_y : 10, \sigma_x : 11$, with order-preserving concatenations of the individual binary strings for multiple qubits. Thus, $\sigma_x \tau_x$ or $\sigma^{(2)}_x \sigma^{(1)}_x$ is assigned [1111] and the binary $Q_{15}$. While the $O_i$ scheme more naturally describes the physics of spin and its coupling to external magnetic fields, the $Q_n$ scheme provides a more natural language for code construction, and it bears a close resemblance to one introduced earlier in a study of $N$ qubits \cite{ref25}. The $Q_n$ scheme is immediately useful when considering the generating operators of spaces over a larger number of qubits, for it  allows easy generalization to such larger number of qubits, each additional qubit enlarging the binary string by two characters 0 and 1. Thus, in a three-qubit register, [000010], would be $\mathcal{I}^{(3)} \mathcal{I}^{(2)} \sigma_y ^{(1)}$, while [110110] would be the operator $\sigma_x ^{(3)} \sigma_z ^{(2)} \sigma_y ^{(1)}$. Reading these numbers as binary representations of ordinary base 10 would allow a labeling also as $Q_n$, the above being $Q_2$ and $Q_{54}$ of the 63 three-qubit operators. Further, the Clifford algebra of multiplication of Pauli matrices translates nicely into the binary addition of the corresponding factors to yield the product to within an irrelevant multiplicative constant. We will make use of both the $O_i$ and the $Q_n$ notations in this paper, using the $O_i$ language when arguments of symmetry become important and using the $Q_n$ language when looking at code design. 

\begin{table*}
\begin{center}
\begin{tabular}{|c|c|c|c|c|c|c|c|c|c|c|c|c|c|c|c|}

\hline
[0001]&[0010]&[0011]&[0100]&[0101]&[0110]&[0111]&[1000]&[1001]&[1010]&[1011]&[1100]&[1101]&[1110]&[1111]\\
\hline
$Q_1$ & $Q_2$ & $Q_3$ & $Q_4$ &  $Q_5$ & $Q_6$ & $Q_7$ &$Q_8$ & $Q_9$ &$Q_{10}$ & $Q_{11}$& $Q_{12}$ &  $Q_{13}$  &  $Q_{14}$&    $Q_{15}$  \\
\hline
$O_3 $&$O_{10} $&$O_9 $&$O_2 $&$O_4 $&$O_{12} $&$O_{11}$ & $O_6$&$O_8$&$O_{14}$&$ O_{16}$&$O_5$&$O_7$&$O_{15}$&$O_{13}$\\ 
\hline 
$\frac{1}{2}\tau_z $&$\frac{1}{2}\tau_y $&$\frac{1}{2}\tau_x $&$\frac{1}{2}\sigma_z$&$\frac{1}{4}\sigma_z\tau_z $&$\frac{1}{4}\sigma_z\tau_y $&$\frac{1}{4}\sigma_z\tau_x $&$\frac{1}{2}\sigma_y $&$\frac{1}{4}\sigma_y\tau_z $&$\frac{1}{4}\sigma_y\tau_y $&$\frac{1}{4}\sigma_y\tau_x$&$\frac{1}{2}\sigma_x $&$\frac{1}{4}\sigma_x\tau_z $&$\frac{1}{4}\sigma_x\tau_y $&$\frac{1}{4}\sigma_x\tau_x $\\
\hline
$-\frac{i}{2}\gamma_1\gamma_2$&$-\frac{i}{2}\gamma_3\gamma_1$&$-\frac{i}{2}\gamma_2\gamma_3$&$\frac{1}{2}\gamma_4$&$\frac{i}{4}\gamma_5\gamma_3$&$\frac{i}{4}\gamma_5\gamma_2$&$\frac{i}{4}\gamma_5\gamma_1$&$-\frac{i}{2}\gamma_5\gamma_4$&$\frac{1}{4}\gamma_3$&$\frac{1}{4}\gamma_2$&$\frac{1}{4}\gamma_1$&$-\frac{1}{2}\gamma_5$&$-\frac{i}{4}\gamma_3\gamma_4$&$-\frac{i}{4}\gamma_2\gamma_4$&$-\frac{i}{4}\gamma_1\gamma_4$\\

\hline
$B_3$&$G_1$&$R_2$&$B_1$&$G_0$&$R_4$&$G_4$&$R_3$&$G_3$&$B_4$&$R_0$&$G_2$&$R_1$&$B_0$&$B_2$\\
\hline
$A^{\flat}$&$A$ & $A^{\sharp}$ & $B^{\flat}$ &  $B$ & $B^{\sharp}$ &$C^{\flat}$ & $C$ &$C^{\sharp}$ & $D^{\flat}$ & $D$& $D^{\sharp}$ &  $E^{\flat}$ & $E$    &   $E^{\sharp}$ \\
\hline
\end{tabular}
\end{center}
\caption{Dictionary for the fifteen operators as binary labels, in the $Q_n$ language, in the $O_i$ language, as direct products of individual Pauli matrices, as Dirac gamma matrices, and in color-flavor and musical notation. The last three occur in Sections IV and V, and the various representations used throughout this paper can be translated with the mapping found here.}
\end{table*}

The two $Q_n$ and $O_i$ languages can be transcribed with the aid of the dictionary in Table I. Along with the $Q_n$ and $O_i$ mappings in Table I, we also list other associations that can be drawn with the generating operators of 2-qubit space. Besides the correspondence to tensors of two Pauli spin matrices initially introduced, we give also a mapping to the the Dirac gamma matrices of relativistic quantum physics \cite{ref26}. These constitute four four-vectors $\gamma_{\mu}$, $\mu =1-4$, denoted V and obeying anti-commutation relations, six anti-symmetric products of two of them denoted T(ensor) as $\sigma_{\mu \nu} = -\frac{i}{2} \gamma_{\mu} \gamma_{\nu}$, a P(seudo-scalar) $\gamma_5$ that is the product of all four gamma matrices, and four pseudo-vectors A given by $i\gamma_5 \gamma_{\mu}$. Since they are widely used throughout particle physics and quantum field theory, the perspective they provide may also be of interest to our discussion of the Kirkman schoolgirls problem. A crucial role in the Kirkman problem is played by a sub-set of seven operators which, in Dirac language, correspond to one A, three V, and three T of the other three indices to the one chosen in A. Other possibilities would have similarly one V plus three each of T and A, or P plus the six T. These correspondences may provide interesting insights into the occurrences of such combinations in relativistic field theories besides our discussion in this paper in terms of two quantum spins.

In addition in Table I, we also include two more mappings that provide an alternative to matrix representations. The first uses language already familiar in particle physics and `chromodynamics' labeling the elements with one of three colors [B, R, G] and with one of five flavors as indices from 0 to 4. The second borrows notation from music theory, assigning labels as one of five notes [A, B, C, D, E] and one of three signatures as sharp, neutral, or flat. We will develop these chromatic and musical representations in the latter half of this paper but list all here in Table I to give a unified reference for the entire paper. This dictionary serves to highlight the range of associations we can draw to the basis set of operators of two qubits. And it is not exhaustive; indeed, we could continue to construct mappings from the basis operators to any collection of distinct elements, such as names of the schoolgirls, points in space, or even distinct tastes and smells!

In the generating group of 2-qubit operators, both under multiplication of any two or upon forming commutators of any pair, the algebraic or group symmetry closes within the set to give one among the sixteen to within numerical factors such as 0 and $\pm i$. We reproduce \cite{ref24} in Table II the commutators of this su(4) Lie algebra because it plays a crucial role in our discussion. A similar table can be made for products (instead of commutators) of pairs of $O_i$ expressed as one among the set of sixteen because of the Pauli-Clifford structure, and we could, of course, render such tables into the $Q_n$ language through the dictionary between $O_i$ and $Q_n$.

\begin{table*}
\begin{center}
\begin{tabular}{|c||c|c|c|c|c|c|c|c|c|c|c|c|c|c|c|c|}

\hline
$O_X $&$O_2 $&$O_3 $&$O_4  $&$O_5  $&$O_6 $&$O_7 $&$O_8 $&$O_9 $&$O_{10} $&$O_{11} $&$O_{12}$&$O_{13}$&$O_{14} $&$O_{15} $&$ O_{16}  $\\ \hline \hline
$O_2 $&$0   $&$0    $&$0    $&$iO_6  $&$-iO_5  $&$iO_8 $&$-iO_7$&$0     $&$0     $&$0     $&$0      $&$iO_{16} $&$-iO_{15}$&$iO_{14}  $&$-iO_{13}  $\\
\hline
$O_3 $&$0   $&$0    $&$ 0    $&$0     $&$0      $&$0     $&$ 0 $&$iO_{10} $&$-iO_9 $&$iO_{12} $&$-iO_{11} $&$iO_{15} $&$-iO_{16}  $&$-iO_{13} $&$ iO_{14} $\\ \hline
$O_4 $&$0   $&$0    $&$0    $&$iO_8  $&$-iO_7  $&$\frac{i}{4}O_6$&$-\frac{i}{4}O_5$&$iO_{12} $&$-iO_{11}$&$\frac{i}{4}O_{10}$&$-\frac{i}{4}O_9$ &$0$&$0$&$0$&$0$\\
\hline
$O_5 $&$-iO_6$&$ 0 $&$ -iO_8 $&$ 0 $&$ iO_2 $&$ 0$&$ iO_4 $&$ 0 $&$ 0 $&$-iO_{16}$&$-iO_{14}$&$ 0 $&$iO_{12}$&$0$&$iO_{11}$\\
\hline
$O_6 $&$iO_5 $&$ 0 $&$iO_7$&$-iO_2$&$0$&$ -iO_4 $&$0$&$0$&$ 0 $&$ iO_{13} $&$iO_{15}$&$-iO_{11}$&$0$&$-iO_{12}$&$ 0 $\\
\hline
$O_7 $&$-iO_8 $&$0$&$-\frac{i}{4}O_6$&$0$&$iO_4$&$0$&$\frac{i}{4}O_2$&$iO_{15}$&$-iO_{13}$&$0$&$0$&$\frac{i}{4}O_{10}$&$0$&$-\frac{i}{4}O_9$&$0$\\ \hline
$O_8 $&$iO_7 $&$ 0 $&$\frac{i}{4}O_5$&$-iO_4$&$ 0 $&$-\frac{i}{4}O_2$&$0$&$iO_{14}$&$-iO_{16}$&$0$&$0$&$0$&$-\frac{i}{4}O_9$&$0$&$\frac{i}{4}O_{10}$\\
\hline
$O_9 $&$0   $&$-iO_{10}$&$-iO_{12}$&$0$&$0$&$-iO_{15}$&$-iO_{14}$&$0$&$iO_3$&$0$&$iO_4$&$0$&$iO_8$&$iO_7$&$0$\\
\hline
$O_{10} $&$0   $&$iO_9$&$iO_{11}$&$0$&$0$&$iO_{13}$&$iO_{16}$&$-iO_3$&$0$&$-iO_4$&$0$&$-iO_7$&$0$&$0$&$-iO_8$\\
\hline
$O_{11}$&$ 0 $&$-iO_{12}$&$-\frac{i}{4}O_{10}$&$iO_{16}$&$-iO_{13}$&$0$&$0$&$0$&$iO_4$&$0$&$\frac{i}{4}O_3$&$\frac{i}{4}O_6$&$0$&$0$&$-\frac{i}{4}O_5$\\
\hline
$O_{12}$&$0$&$iO_{11}$&$\frac{i}{4}O_9$&$iO_{14}$&$-iO_{15}$&$0$&$0$&$-iO_4$&$0$&$-\frac{i}{4}O_3$&$0$&$0$&$-\frac{i}{4}O_5$&$\frac{i}{4}O_6$&$0$\\
\hline
$O_{13}$&$-iO_{16}$&$-iO_{15}$&$0$&$0$&$iO_{11}$&$-\frac{i}{4}O_{10}$&$0$&$0$&$iO_7$&$-\frac{i}{4}O_6$&$0$&$0$&$0$&$\frac{i}{4}O_3$&$\frac{i}{4}O_2$\\
\hline
$O_{14}$&$iO_{15}$&$iO_{16}$&$0$&$-iO_{12}$&$0$&$0$&$\frac{i}{4}O_9$&$-iO_8$&$0$&$0$&$\frac{i}{4}O_5$&$0$&$0$&$-\frac
{i}{4}O_2$&$-\frac{i}{4}O_3$\\
\hline
$O_{15}$&$-iO_{14}$&$iO_{13}$&$0$&$0$&$iO_{12}$&$\frac{i}{4}O_9$&$0$&$-iO_7$&$0$&$0$&$-\frac{i}{4}O_6$&$-\frac{i}{4}O_3$&$\frac{i}{4}O_2$&$0$&$0$\\
\hline
$O_{16}$&$iO_{13}$&$-iO_{14}$&$0$&$-iO_{11}$&$0$&$0$&$-\frac{i}{4}O_{10}$&$0$&$iO_8$&$\frac{i}{4}O_5$&$0$&$-\frac{i}{4}O_2$&$\frac{i}{4}O_3$&$0$&$0$\\
\hline
\end{tabular}
\end{center}
\caption{Table of commutators. With operators $O_i$ in the first column and $O_j$ in the top row, each entry provides the commutator $[O_i,O_j]$.}
\end{table*}

Among the many interesting patterns that can be discerned from Table II are sub-group or sub-algebras, the most important for our purposes in this paper of the Kirkman problem being one of seven $O_i$ that close under commutation. (Other sub-groups such as the orthogonal group of rotations in five dimensions \cite{ref11} will not be discussed here.) A glance at the table reveals seven zeroes in each row, that is, that for every $O_i$, there are six others besides itself with which it commutes, and those six close among themselves under commutation. This is, therefore, a su(2) $\otimes$ u(1) $\otimes$ su(2) sub-algebra involving only seven among the set of fifteen $O_i$, and there is one such sub-algebra for each $O_i$. This feature has already been exploited for some Hamiltonians of interest in molecular physics and quantum optics \cite{ref24} and for some two qubit states \cite{ref19}, and holds the key to our resolving of Kirkman designs. A convenient rendering of the set of seven is shown in Fig. 1, a diagram called the Fano Plane \cite{ref5}.

% === Fano with Binary Coordinates ======================================================
\captionsetup{justification=centering}
\begin{figure}
\scalebox{1.0}{\includegraphics[width=3in]{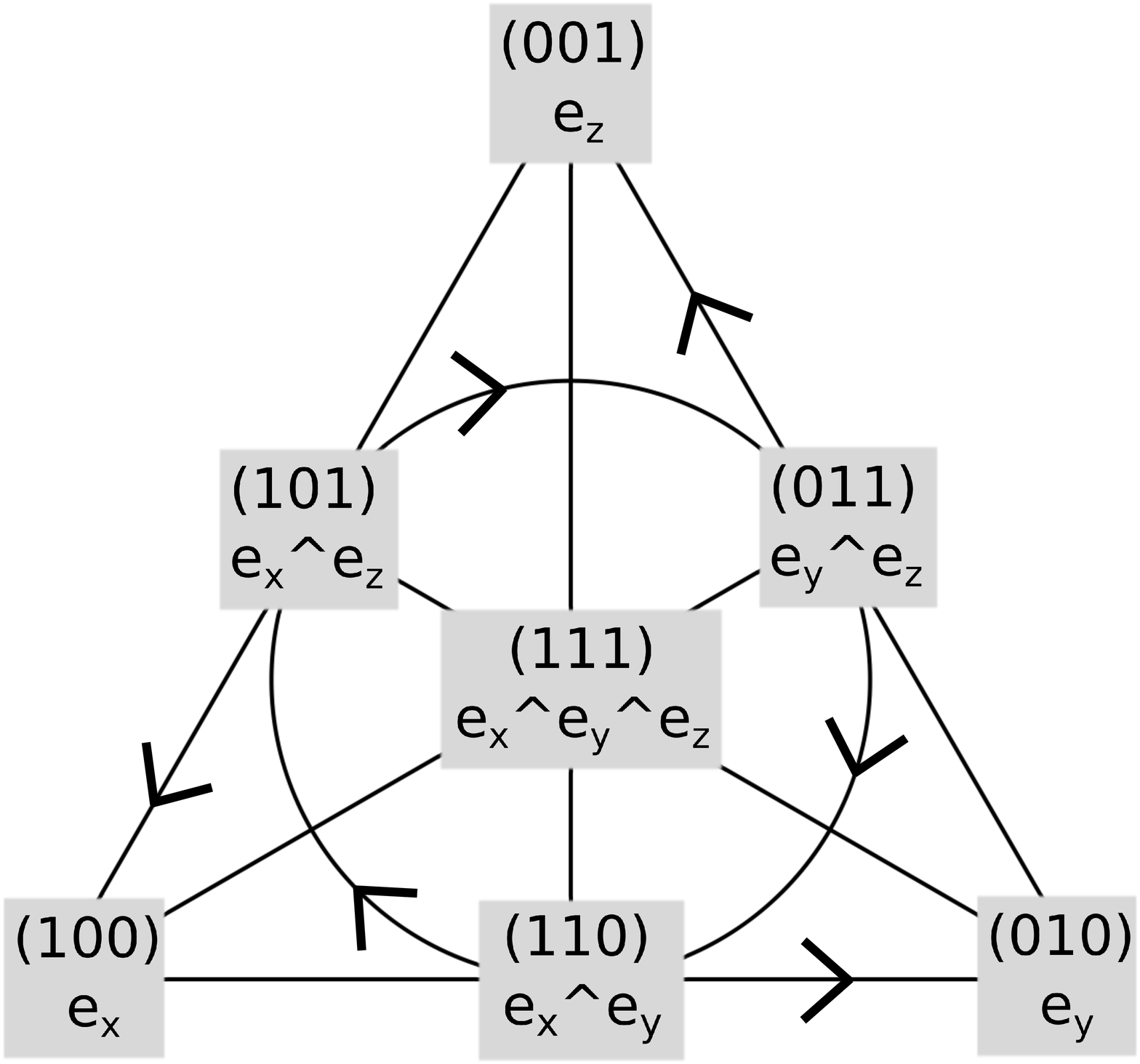}}
%\fig[height=2.8in]{Images/Fig1.eps}{\textit
\caption{The Fano Plane of projective geometry. At each vertex sits a binary coordinate with one entry non-zero and a corresponding geometric base. Outer products of the vertices define the higher grade points with two unit entries. The Fano Plane is the geometry of the (7, 3, 1) design.}
\end{figure}
% ==============================================================================================================
Fig. 1 of seven points lying on seven lines, with each point ``incident" \cite{ref8} on three lines and each line defined by three points, is a central figure of finite projective geometries and is called the Fano Plane \cite{ref5}; indeed, it is the smallest projective plane and is denoted by PG(2,2) \cite{ref8,ref15}. Our interest in this paper is the aspect of no two points sharing more than one line in common, precisely as required for the Kirkman arrangement of triples (the lines) with no pairs in common. Every pair of $O_i$ defines uniquely the third to lie on one of the seven lines. The central element, the center of the triangle, commutes with all the other six elements on the vertices, whereas the other four lines obey the usual Lie commutations of quantum angular momentum, the sign specified by the direction of the arrow. Note that the inscribed circle is also considered a fourth line on par with the three sides of the equilateral triangle, no distinction made in projective geometries between such a ``line at infinity through three points at infinity" and lines through points at finite locations \cite{ref11,ref19}. The unarrowed medians may be referred to as ``commuting," and the other four arrowed lines of Fig. 1 as ``cyclic," triplets. In their Clifford structure, the product of two $O_i$ gives the third, in any order, among the commuting triplets, whereas such products in the cyclic triplets differ by a plus/minus in the order along/against the arrow direction. Any $O_i$ or $Q_n$ can sit at the center (111) with the six other zeroes of Table II providing the other points of the triangle.

\section{Constructing Kirkman arrangements}

The groups and symmetry algebras of the previous section, especially the su(2) $\otimes$ u(1) $\otimes$ su(2) sub-algebra of su(4), are the basis of our construction of Kirkman arrangements. Other arguments of combinatorics have been used \cite{ref12,ref13} but we exploit the identification of 15 with the generators of two-qubit systems for this purpose. Before proceeding to supplementing the triangle in Fig. 1 with a cube for the Kirkman arrangement of 15 in 5 rows of 3 for 7 days, that triangle of the Fano Projective Plane with 7 points on 7 lines can itself be rendered in the language of Kirkman as a simplified arrangement of just 7 girls in a single row of three for each of 7 days, each girl to be in three days and with each of her other friends once and only once. The seven lines of the diagram provide the required 7 triplets. In alternative languages, this projective geometry PG(2, 2) is the Steiner triple 2-(7, 3, 1), whereas Kirkman's full arrangement is PG(3, 2) or the Steiner 2-(15, 3, 1) \cite{ref8,ref11,ref15}. The mathematical literature draws a distinction between ``Steiner Triple Systems (STS)," $(v, 3, 1)$, and ``Kirkman Triple Systems (KTS)," $(3n, 3, 1)$, getting a Kirkman arrangement termed as ``resolvable" only when $n$ is odd \cite{ref12}. An even simpler example than the Kirkman schoolgirls' $n=5$ is for $n=3$ of $v=9$ schoolgirls with $b=12$ triplets arranged in 3 rows of $k=3$ each for $r=4$ days, or the trivial $n=1$ of 3 schoolgirls in one row for one day. Standard terminology uses $v$ for ``varieties" and $b$ for ``blocks". In the language of projective geometry, they stand, respectively, for points and lines, and a ``duality" obtains between them \cite{ref8,ref12}. Since two points define uniquely a line, that no pairs recur is automatic in projective geometry, equally so in our usage of Lie-Clifford algebra in the qubit language. The trivial 2-(3, 3, 1) design expresses that the product of any two $O_i$ or $Q_n$ gives uniquely a third, either as a commuting or a cyclic triplet. 

Starting with the same seven triplets/lines, assigning one to each of the seven days of the week, the other 28 required triplets to fill all five rows for each day in a full Kirkman arrangement are conveniently rendered in terms of a cube to accompany the triangle in Fig. 1, a cube's edges, face and body diagonals also numbering 28 in all. This combination also occurs in \cite{ref20} for a different purpose. As stated above, every $O_i$ in Table II provides a set of seven points/girls corresponding to the zeroes in that table. The other eight non-zero entries provide the corners of the cube. Picking any one of them as the starting corner, lowest left in Fig. 2, the other seven fall into place as follows. From that corner are three edges, three faces, and the body of the cube, seven in all and we place the seven from the triangle as the corresponding mid-points, that is, mid-point of the edges, face-centers of the faces and body-center of the cube. The corresponding corner at the other end is then uniquely the third member of the commutator triplet with the previous two, all again read off Table II.

Some notation is helpful at this point. As already indicated in Fig. 1, a natural set of three binary bits labels the seven points, a triplet of 0/1 assigned as shown for seven possibilities. Thinking geometrically, we have $(x, y, z)$, with $x=0$ the right edge, $y=0$ the left edge, and $z=0$ the bottom edge of the triangle, ``excitation" off any edge being 1, or equivalently, in colors (B, G, R) for the terminology used in Sec. IV and the seven colors of the rainbow. Each binary bit string can be reinterpreted as a basis in a grade-3 geometric algebra, particularly useful when considering algorithm construction. An alternative to picking one $O_i$ for the center and referring to the edges is to view Fig. 1 from the perspective of projective geometry's duality between points and lines by assigning to the vertices of the triangle the numbers (001), (010) and (100), the other points then following by binary addition modulo 2 along any line or from the commutator Table II. The points on the cube in Fig. 2 are similarly rendered in a four-bit binary scheme. To avoid confusion with the earlier binary notation for $Q_n$ and shown in square brackets in Table I, we now use round brackets that apply to Figs. 1-3 and Table III in this section. Together, the round brackets for labeling the points in the figures and square brackets for the $Q_n$ that occupy them give the Kirkman designs we will generate. 

To extend the binary notation to the cube, we supplement with a new first binary entry (say ``time" t for the space-time language or U for ultraviolet in the four-color scheme), denote one chosen corner as (1000), make each of the previous seven in the triangle into a quartet with initial entry 0, that is, $(0xyz)$, and place them as the mid-points referred to in the previous paragraph. The seven other corners of the cube are then obtained through them from the chosen (1000) by binary addition. These corners, therefore, have one unit of excitation in that first (or U) entry, as shown in Fig. 2, besides the excitations in $(x, y, z)$ and (B, G, R) already in the seven points of Fig. 1. All 15 schoolgirls or $O_i$ operators are thus uniquely specified in binary or color scheme. Specific renderings will be presented in later sections. We will also carefully distinguish between these specific renderings with two differing binary labels, using square brackets for $O_i$ or $Q_i$ as in the first row of Table I, while reserving round brackets for the points on the triangle and cube of Figs. 1 and 2 (and a tetrahedron to be introduced below) that they occupy. 

% === Cube & Tetrahedron =============================================
%\begin{minipage}[c]{0.5\textwidth}
% \fig[height=3in]{Images/Fig2.eps}{\textit
\begin{figure}
\scalebox{1.0}{\includegraphics[width=3in]{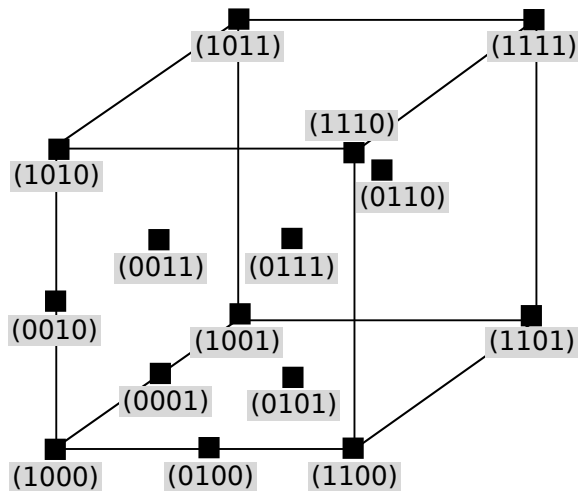}}
\caption{Fifteen binary labels on the cube. For clarity, repeated points of the mid-points and face elements are not shown, only the left (0011), bottom (0101), and back (0110) face-centers. The body-center is (0111). The cube, along with the Fano Plane, provide one geometry for the (15,3,1) design.}
\end{figure}
%\end{minipage}
%\begin{minipage}[c]{0.5\textwidth}
 %\fig[height=3in]{Images/Fig3.eps}{\textit
\begin{figure}
\scalebox{1.0}{\includegraphics[width=3in]{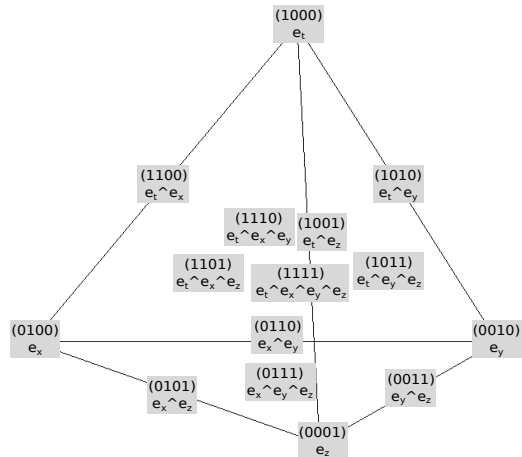}}
\caption{Binary labels of geometric bases in grade-4 space. The tetrahedron contains fifteen geometric bases that can be labeled with 4-bit binary strings. The tetrahedron provides another geometry for the (15,3,1) design.}
\end{figure} 
%\end{minipage}
% ====================================================================

This aspect of a binary notation is a natural way of adding a dimension \cite{ref27} or color. Indeed, the triangle in Fig. 1 is already such a generalization of just two colors with only one triplet, (01), (10), and (11) of the right edge, supplemented now by the third index of the new vertex (100) on the left; its links to the previous three provides seven points in all - seven colors of the rainbow from three primary colors. This also points to an alternative geometrical picture than the triangle plus cube. Just as one goes from one line, the right edge, to the two-dimensional triangle by introducing a new vertex, a similar step into four bits or colors is achieved by a new vertex and going from equilateral triangle to a tetrahedron with four such faces. This is shown in Fig. 3 along with the same natural binary labeling from (0001) to (1111) that extends the base plane $(0xyz)$ through the introduction of the new vertex (1000). This natural nesting of ``simplexes" \cite{ref28} also corresponds to the projective geometry extensions of such finite geometries of 3, 7, 15, $\ldots$, $2^{n+1}-1, \ldots$ elements as PG(1,2), PG(2,2), PG(3,2), $\ldots$, PG($n$,2), $\ldots$. Indeed, \cite{ref13} notes that Davis had already in 1897 used a cube's 8 vertices, 6 faces, and body as a whole as simplicial elements. In such a picture, the Fano Plane PG(2,2) in Fig. 1 is already viewed similarly as 3 vertices (0-simplex), 3 edges (1-simplex) and triangle (2-simplex), and the generalization to the PG(3,2) tetrahedron as 4 vertices, 6 edges, 4 faces (2-simplex) and tetrahedron (3-simplex) to cover the 15 points.

Of use also for an algorithm to generate Kirkman arrangements that we will discuss below, the binary and simplex notation cast the problem in terms of four base vectors $(e_1, e_2, e_3, e_4)$, six bi-vector lines (1-simplex) $(e_1 \wedge e_2$, etc.), four tri-vector faces (2-simplex) and the one quadravector (3-simplex) of the tetrahedron itself. These are shown in Figs. 1 and 3. With four $Q_n$ as a seed, the algorithm will generate a full KTS. 

The next step toward the full Kirkman arrangement is to identify each day of the week also in binary notation. A natural choice is (001) for MON(day), (010) for TU(esday), (011) for WED(nesday), (100) for TH(ursday), (101) for FRI(day), (110) for SAT(urday) and (111) for SUN(day). The assignment of the first row of three for each day places the similarly labeled point of the triangle as the middle member with the other two of the triplet as its partners. The four other points of the triangle left out are then placed as the middle member of the other four rows for each day and the partners filled in as the partners for the Lie-Clifford triplets from the cube. Thus, for a Kirkman arrangement, the triangle provides one triplet for each day of the week and the cube the remaining 28 triplets. All but the central element of the triangle have repeats in the cube with its sets of three equivalent opposite faces and four parallel edges along the $x$, $y$ and $z$ directions of its three dimensions. These could be denoted by primes for the other mid-points of edges and faces but are not shown in Fig. 2. 

All 35 triplets of triangle plus cube are thus accommodated to give a pattern satisfying, by construction, the requirements of the Woolhouse-Kirkman problem. Of these triplets, 15 are commuting and 20 cyclic. For any $O_i$, the six zeroes in that row of Table 1 form 3 pairs that give the commuting lines (medians) in Fig. 1, with that particular $O_i$ standing at the center. The other eight non-zero entries along that row provide, in 4 pairs, the corners of the cube in Fig. 2.

As noted, a three-dimensional tetrahedron as shown in Fig. 3 provides an alternative to the triangle + cube picture. The fifteen points or schoolgirls are, of course, the four vertices, mid-points of six edges, four face-centers, and one body-center. The 35 triplets can be grouped into six classes: six edges, twelve medians of three on each face, four circles on each face, four body altitudes, three connecting the body center to two mid-points of edges, and six connecting two face centers to a mid-point of an edge. A similar classification of six categories is in \cite{ref13}. Each of the four faces of the tetrahedron constitutes a Fano Plane as in Fig. 1 which forms the base of the tetrahedron in Fig. 3.

Consider the example of $O_4 = \sigma_z \tau_z/4$ as the commuting element, occupying the central point (111) of Fig. 1. The three medians are symmetrically on par and their ends can be occupied by any of the three commuting pairs, say, $(001) : O_{16} = \sigma_y \tau_x/4, (110) : O_{15} = \sigma_x \tau_y/4; (010) : O_2 =\sigma_z/2, (101) : O_3 = \tau_z/2; (100) : O_{14} = \sigma_y \tau_y/4, (011) : O_{13} =\sigma_x \tau_x/4 $. The three pairs are separated by semi-colons and shuffling between them is simply accommodated by rotating the triangle through 60 degrees. Next, to assign entries in the cube, with $O_4$ again in the body center, now (0111) with the first 0 entry, the 4 cyclic pairs can be placed at opposite corners of the cube, again any shuffling between them of little consequence, accommodated by the symmetries of the cube. Thus, choosing one as the bottom left corner shown, we have $(1000) : O_5 =\sigma_x/2, (1111) : O_8 =\sigma_y \tau_z/4; (1100) : O_{12} =\sigma_z \tau_y/4, (1011) : O_9 =\tau_x/2; (1001) : O_{11} = \sigma_z \tau_x/4, (1110) : O_{10} =\tau_y/2; (1101) : O_7 =\sigma_x \tau_z/4, (1010) : O_6 =\sigma_y/2$. From the (1000) left corner, the mid-points of the seven lines from it to the other corners above are the seven points of the triangle with an extra zero for the first number in the quartet labelling: $(0xyz)$, with the $x$-axis the horizontal from left to right, the $y$-axis the vertical, and the $z$-axis going from the left corner into the page. There are three edge, three face-diagonal, and one body-diagonal lines from the left corner. The middle three are commuting, the other four cyclic Lie-Clifford triplets. On any line, the sum of two modulo 2 gives the third; alternatively, the sum of all three modulo-2 reduces all entries to zero. The point-mappings of this design are given in Table III with $O_4$ (or $Q_5$) chosen as the center of Figs. 1 and 2, and $O_5$ (or $Q_{12}$) as the bottom front-corner of the cube or top vertex of the tetrahedron in Fig. 3 with, correspondingly, $O_8$ (or $Q_9$) the body-center. Note again that the modulo 2 binary addition and numbers in the figures and in this paragraph are set within round brackets to distinguish from the square ones shown for $Q_n$ in the top row of Table I. 

\begin{table*}
\begin{center}
\begin{tabular}{|l|l|}
\hline
	{(000)} : $O_1,Q_0,X,X$ & (1000) : $O_5,Q_{12},G_2,D^{\sharp}$ \\
	{(001)} : $O_{16},Q_{11},R_0,D$  & (1001) : $O_{11},Q_7,G_4,C^{\flat}$ \\
	{(110)} : $O_{15},Q_{14},B_0,E$ & (1110) : $O_{10},Q_2,G_1,A$ \\
	{(011)} : $O_{13},Q_{15},B_2,E^{\sharp}$ & (1011) : $O_9,Q_3,R_2,A^{\sharp}$ \\
	{(010)} : $O_2,Q_4,B_1,B^{\sharp}$ & (1010) : $O_6,Q_8,R_3,C$ \\
	{(101)} : $O_3,Q_1,B_3,A^{\flat}$ & (1101) : $O_7,Q_{13},R_1,E^{\flat}$ \\
	{(100)} : $O_{14},Q_{10},B_4,D^{\flat}$ & (1100) : $O_{12},Q_6,R_4,B^{\sharp}$\\
	{(111)} : $O_4,Q_5,G_0,B$ & (1111) : $O_8,Q_9,G_3,C^{\sharp}$\\
\hline

\end{tabular}
\end{center}
\caption{ \textit{Mapping of the $D(15,3,1)|Q_{12},Q_{10},Q_4,Q_{11}\rangle$ and the embedded $D(7,3,1)|Q_{10},Q_4,Q_{11}\rangle$ design}. On the left sits the coordinates of the points' geometric bases and on the right sits the names of the operators in the various representations. We see that the Fano sub-design of the tetrahedron design can be immediately found by looking at the points on the tetrahedron with $t=0$ such that there is a 0 in the first place of the point's binary string. These seven elements in the left column of the table sit as the base of the pyramid in Fig. 3. }
\end{table*}

The convenient binary notation of doublet, triplet, quartet introduced in this section can also be generalized to other place values than the 2 discussed, wherein each step to the left increases by a further power of 2. Thereby, we can envisage variants on the Kirkman triplets or rows of three for other multiples $k$. The simplest such alternative, with place value 1 so that each step and all powers of 1 remain 1, would have (1) representing the number 1, (11) representing 2, (111) representing 3, etc. In this sequence, with $k$ now 2, that is, pairs and not triplets, starting with a right edge connecting (01) and ((11), the next step to a new vertex (111) that with the previous (001) and (011) makes a triangle, and the next of new vertex (1111) making a tetrahedron, would represent PG(3,1) of 4 points/girls pairing in two rows for each of three days, with all six pairs/edges distinct. More generally, for place value $q$, PG(3,$q$) has $(q^2+1)(q+1)$ girls walking in $(q^2+1)$ rows of $(q+1)$ each for $(q^2+q+1)$ days. The number of lines/blocks $(q^2+1)(q^2+q+1)$ is said to be partitioned into ``spreads"/days, such a partition called a ``packing" \cite{ref12} or a ``parade" \cite{ref14} of such a generalized Kirkman arrangement. For the $q=k-1=2$ Kirkman problem, \cite{ref13} provides a nice history of the problem and notes that there are 80 non-isomorphic STS but only 4 KTS \cite{ref14}, isomorphism referring to interchange by duality of the projective space. Our placement of the seven triplets from the triangle in Fig. 1 in the top row of Table IV, one for each day/spread is an aspect of the set of lines of PG(2,2) inside PG(3,2) necessarily lying in different spreads as noted in \cite{ref12}. 

\subsection{An algorithm to generate Steiner $(v, 3, 1)$ and Kirkman $(3n, 3, 1)$ designs}

To generate designs $(v, k, \lambda)$, with $k=3, \lambda =1$, we have set up a C++ application that generates design systems and corresponding color and sound designs \cite{ref29}. Sets of multi-dimensional arrays describe the composition of the design system. The computer first initializes the space of points of the design with identity elements and then demands input from the user. The user's input defines the initial conditions or the ``seed" of the design system. Once the initial conditions have been given, the computer performs our algorithm until a mapping of operators to points that fulfills the initial conditions provided by the user has been found. 

For example, consider the 0-simplex or `point' design $D(1,1,1)$, with design characteristics that mirror the $(v,k,\lambda)$ description. The 0-simplex design contains 1 point to which any of our operators $Q_i$ is mapped. The point functions as the container, which can be represented as a binary coordinate, a blade in a geometric space $\hat{e}_i$, or an abstract ket $|n\rangle$, and the operator functions as the design-element, which can be represented as a matrix of complex values, a colored stain, a musical note, or a schoolgirl's name. The action of the algorithm is to map design-elements to specific points. The $D(1,1,1)$ takes just one initial condition, which we represent in quantum-physical Dirac notation as $|Q_i\rangle$. The action of generating the 0-simplex design is described by the operation $D(1,1,1)|Q_i\rangle$. It outputs, of course, one $Q_i$ at one point. For generalization to more non-trivial steps, the actual operation is represented by $D(1,1,1)|Q_i\rangle \equiv |0\rangle + Q_i|1\rangle$ and with this the only operation for this trivial point design, the output is also $|0\rangle + Q_i|1\rangle$. Dropping the initial null output, the result is $Q_i|1\rangle$ and read as a point labelled (1) and $Q_i$ placed on it. Any of the fifteen choices of $Q$ may be placed at a point. 

Next, a line or 1-simplex design clearly requires two points as the seed, and it generates the simplest KTS, in design notation written $D(3,3,1)|Q_i, Q_j\rangle$, which outputs $v=3$ points that each contain a unique object, whether a schoolgirl or a quantum operator. The two seed objects/operators lie at the points within the Dirac ket as given by the initial conditions. As in the previous paragraph, this is opened as 

\begin{equation}
(|0\rangle+Q_i|1\rangle) \otimes (|0\rangle+Q_j|1\rangle) = |00\rangle+Q_i|10\rangle+Q_j|01\rangle+Q_iQ_j|11\rangle,
\end{equation}
and again, after dropping the initial null state, three points (10), (01), and (11) laid on a line with correspondingly $Q_i$, $Q_j$ and their product placed on them. Again, as in the previous paragraph, any pair of $Q$ may be chosen without restriction because a unique third results from their Clifford multiplication, that multiplication carried out by binary addition of their square bracket 4-bit binary in Table I. That $Q$ is placed at the point (11) as a result of this operation $D(3,3,1)|Q_i, Q_j\rangle$. Any of the three edges of Fig. 1 provides such an example upon dropping a common 0 to reduce them to a 2-bit binary. For purposes of extension to higher designs below, note that the two seeds may be regarded as the end-points, vertices, of the line generated by $D(3,3,1)$, with its output as the mid-point of that line. 

The next Fano design (7, 3, 1) requires three seeds written as $D(7,3,1)|Q_i, Q_j, Q_k\rangle$. The number of seeds is always the base-2 logarithm of $v+1$. A direct set of seven operators comes from our earlier observation of seven zeroes in each row of Table I. We can construct such a set of seven operators by mapping the initial conditions to the grade-1 outer vertex points (100), (010), and (001) as in Fig. 1 so that we now write

\begin{equation}
(|0\rangle + Q_i|1\rangle) \otimes (|0\rangle + Q_j|1\rangle) \otimes (|0\rangle +Q_k|1\rangle)  
\end{equation}
which opens out to 

\begin{equation}
|000\rangle+Q_i|100\rangle+Q_j|010\rangle+Q_k|001\rangle+Q_iQ_j|110\rangle+Q_iQ_k|101\rangle+Q_jQ_k|011\rangle+Q_iQ_jQ_k|111\rangle.
\end{equation}
Again, discarding the initial null ket, this generates the triangle in Fig. 1 with $Q$ occupying the seven points starting with three seeds. Unlike the previous examples with just one or two seeds, there is a restriction on how to choose the seeds now. We must pick seed operators such that the product of two does not return the third seed in the set, i.e., the three seed operators must not themselves form a commuting or cyclic triplet but must be ``linearly independent." The higher grade points are then determined by multiplication of the operators that sit at the grade-1 points with the familiar prescription that the product of two operators on a line returns the third. For example, the point (101) lies on a common line with the points (100) and (001), so the operator that sits at (101) is given by the product of the seed operators $Q_i$ and $Q_k$. The center point (111) is found by multiplying all three operators at the grade-1 points, and this operator will, of course, commute with all other six operators in the design just as we saw in Table II. Alternatively, we could assign one seed to the point (111) and find the design given the other two initial conditions, which gives an useful method to characterize a design by its commuting element. As another alternative view of Fig. 1, starting from two points of the right edge with initial entry zero in $(xyz)$, any other point with $x=1$ will generate the whole triangle. The choice of the third point on that same edge with $x=0$ will, however, only give the (3, 3, 1) design that is the right edge; it takes one of the other four points to give the (7, 3, 1) design. Our prescription of considering the seeds as the vertices ensures generation of the full design. The six other zeroes in Table II for any $O_i$ enters here in making the choice of three seeds from among that set of seven to get the whole pattern.

With this, it is clear how our constructive algorithm generalizes, the next (15, 3, 1) 3-simplex tetrahedral Kirkman design now designated $D(15, 3, 1)|Q_i, Q_j, Q_k, Q_l\rangle$ needing a seed of four $Q$, three from a Fano Plane and one off that plane. That is, starting with any $Q_i$ and two others of the zeros in that row of Table I that generate the Fano triangle, pick any one of the other eight non-zero entries as the fourth to generate the Kirkman design. In binary notation, $(txyz)$, any of these with $t=1$ supplements the other three with $t=0$ to provide a tetrahedron as in Fig. 3. The four seeds, each with one non-zero entry of unity serve as vertices of the tetrahedron. 

The design notation $D(v,k,\lambda)|\mathbf{Q}\rangle$ introduced here gives a means of maximally labeling unique design systems with a minimal number of initial conditions. For most purposes, we map the array of initial conditions $\mathbf{Q}$ to the grade-1 points in the geometric space. In the Fano $(7,3,1)$ design, we found that three initial conditions maximally characterize a design, so we map the three initial conditions to the three grade-1 points of grade-3 space. Similarly for the tetrahedron $(15,3,1)$ design, four initial conditions maximally characterize the design system, so we map to four grade-1 points in a grade-4 space. In the initial array $\mathbf{Q}$, the location of the seed operator in the array corresponds to the binary representation of the point to which the operator is mapped. For example, in a $D(15, 3, 1)|Q_i, Q_j, Q_k, Q_l\rangle$ design, $Q_l$ sits at the point (0001). Our dual notation of square and round brackets proves helpful in this algorithmic construction, the underlying geometrical structures of Figs. 1-3 themselves a result of carrying out the $D$ operation, with $D(v, 3, 1)$ giving line, triangle, and tetrahedron for $v=3, 7, 15$. Specific objects to be placed at the points can vary with different inputs from Table I for $O$ or $Q$, with the Lie-Clifford physics and mathematics determining outputs, all labeled by square brackets.  Thus, our previously developed Table III in this language gives the $D(7,3,1)|Q_{10},Q_4,Q_{11}\rangle$ and the $D(15,3,1)|Q_{12},Q_{10},Q_4,Q_{11}\rangle$ designs. We will use these two specific solutions as examples throughout the rest of this paper but must remember that other designs can also be found if we provide alternative initial conditions; for each $O_i$ or $Q_n$, we have a Table III. 

\section{A four-color scheme}

Our acquaintance with color goes back to our very beginnings, as also the recognition of three basic or primary colors. ``Additive" and ``subtractive" combinations of them have been recognized in the palette of rainbows and paintings for centuries before the scientific understanding through the electromagnetic spectrum and the three types of color pigments in retinal cones. That understanding in terms of a narrow band of wavelengths that our retinas sense, with three types of cells of peak sensitivity in three separate regions between 400 and 700 nm, is only a very small part of color perception itself, which involves also intensities, contrast with other colors in the surround, neural psychology, and even cultural elements. However, the idea of three primary colors and together a colorless ``white," reversing Newton's discovery of splitting the Sun's white light into component colors, has become a paradigm in language and literature for more general synthesis. In the last fifty years, it has become a key element in the study of fundamental particles in physics, with quarks coming in three colors and only colorless combinations (either all three quarks with one of each color or a quark with its anti-quark, ascribed an ``anti-color") realized in the standard model. That model also invokes another degree of freedom, named with another colloquial term ``flavor," there being six flavors of quarks for each color. 

As stated above, a 4-string binary with associated four chroma [U, B, G, R] could give a unique label for each of the 15 entities. Though a parrot's sensory organs might be able to distinguish different 4-chroma colors, the authors or readers/viewers of this paper can only distinguish 3-chroma colors. Therefore, using only 3 basic chroma, but adopting the particle physics language now with five flavors of each, we use such a color-flavor scheme as a natural and convenient language for labeling the Woolhouse-Kirkman girls or the quantum qubit pairs. In this, we exploit the color sensitivity of the human eye to distinguish a range of hues within each primary color, indexing each of [B, G, R] by a hue/flavor index of 0-4.
     
Continuing with our running example of the $D(7,3,1)|Q_{10},Q_4,Q_{11}\rangle$ and the $D(15,3,1)|Q_{12},Q_{10},Q_4,Q_{11}\rangle$ designs, we use the color-flavor mapping of Table I to assign colors to the points of a design system. Thus Fig. 4 is our particular solution of the $(7,3,1)$ design of Fig. 1, with the seven points now also labeled with colors [B, G, R] and subscript flavor indices from 0 to 4. The center of the triangle has been labelled G$_0$ and one of the medians, a commuting line, has its ends labeled with partner flavor 0 of B$_0$ and R$_0$. Taking one of the colors, say B, the other four vertices are ascribed the flavors B$_{1-4}$. As per the seventh paragraph of Sec. III, with the same binary triplet notation for the days of the week, the corresponding point, placed as the middle entry, provides the top triplet row in Table III's Kirkman arrangement. In rendering flavors visually, one possible prescription is provided by visual complex analysis \cite{ref30}. To accommodate fifteen distinct color-flavor pixels, the Hue Saturation Value (HSV) color system with complex numbers is used, color associated with the phase value ($0^{\circ}$ for R, $120^{\circ}$ for B, $240^{\circ}$ for G) and flavor with the norm of the complex number, with flavor 0 the least intense, and subsequent flavors increasing in the intensity or quantity of white in the color. After some experimentation for the most pleasing appearance, we used a roughly exponential scale to model flavor, such that the intensity difference between the lower flavor colors are lower than for the higher flavor quanta. Each of the 15 schoolgirls or equivalent objects is thus uniquely distinguished and Fig. 4 is an example for the seven that we have used as an example throughout.  

Similarly, Fig. 5 is a re-rendering of the cube in Fig. 2, now complete with color and flavor labels as well. Placing G$_0$ again as the body center and labeling the lower left corner (1000) as G$_2$, the other six points of Fig. 4 as mid-points of the edges and facial medians from that left corner, and the other seven vertices of the cube are shown. The vertices of the cube are the G$_{1-4}$ and R$_{1-4}$ that complement the flavors of B already assigned in Fig. 4 and as per the same HSV scheme. Equivalency of opposite faces of a cube is reflected in double repetition of B$_0$, B$_2$, and B$_3$, and equivalency of horizontal and vertical lines in triple repetition of R$_0$, B$_1$, and B$_4$ and, for convenience, are shown by primes, and double and triple primes in Table IV. In filling the remaining four rows of the Kirkman arrangement in Table IV with the 28 triplets from the cube in Fig. 5, the following proves convenient. For each day of the week, place one each of the other four points of Fig. 4 not in the first row as the middle entries for the lower four rows, and complete the triplet as given by Fig. 5. Within each day, of course, the rows can be shuffled around without consequence for the Kirkman arrangement.

% === Fano, Cube, and Tetrahedron with Colors ======================================================
%\captionsetup{justification=centering}
%\begin{minipage}[c]{0.3\textwidth}
 %\fig[height=1.8in]{Images/Fig4.eps}{\textit
\begin{figure}
\scalebox{1.0}{\includegraphics[width=3in]{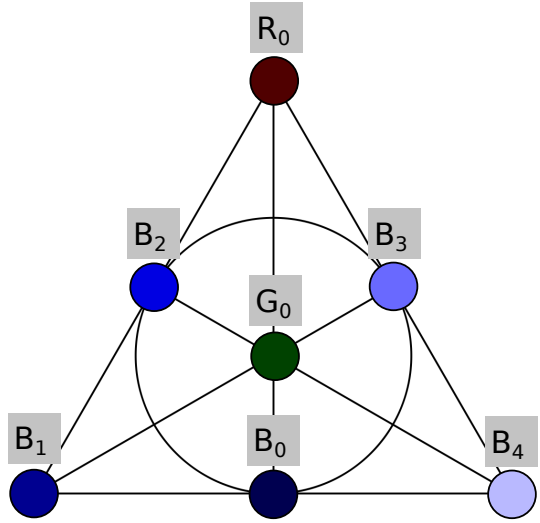}}
\caption{A Fano color design. The elements in this KTS represent the solution to the design $D(7,3,1)|Q_{10},Q_4,Q_{11}\rangle$.}
\end{figure}
%\end{minipage}
%\begin{minipage}[c]{0.3\textwidth}
 %\fig[height=1.8in]{Images/Fig5.eps}{\textit
\begin{figure}
\scalebox{1.0}{\includegraphics[width=3in]{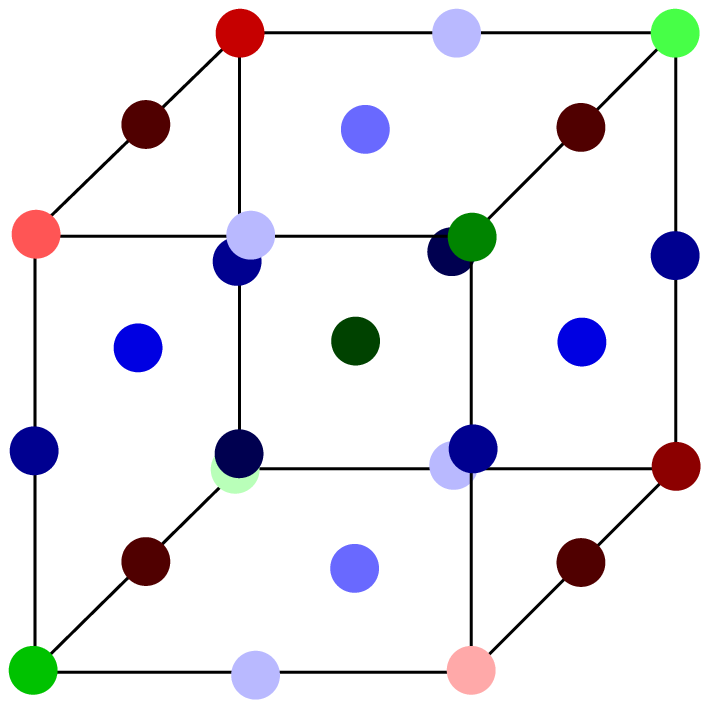}}
\caption{A cubic color design. The elements in this KTS represent the solution to the design $D(15,3,1)|Q_{12},Q_{10},Q_4,Q_{11}\rangle$.}
\end{figure}
%\end{minipage}
%\begin{minipage}[c]{0.3\textwidth}
%\fig[height=1.8in]{Images/Fig6.eps}{\textit
\begin{figure}
\scalebox{1.0}{\includegraphics[width=3in]{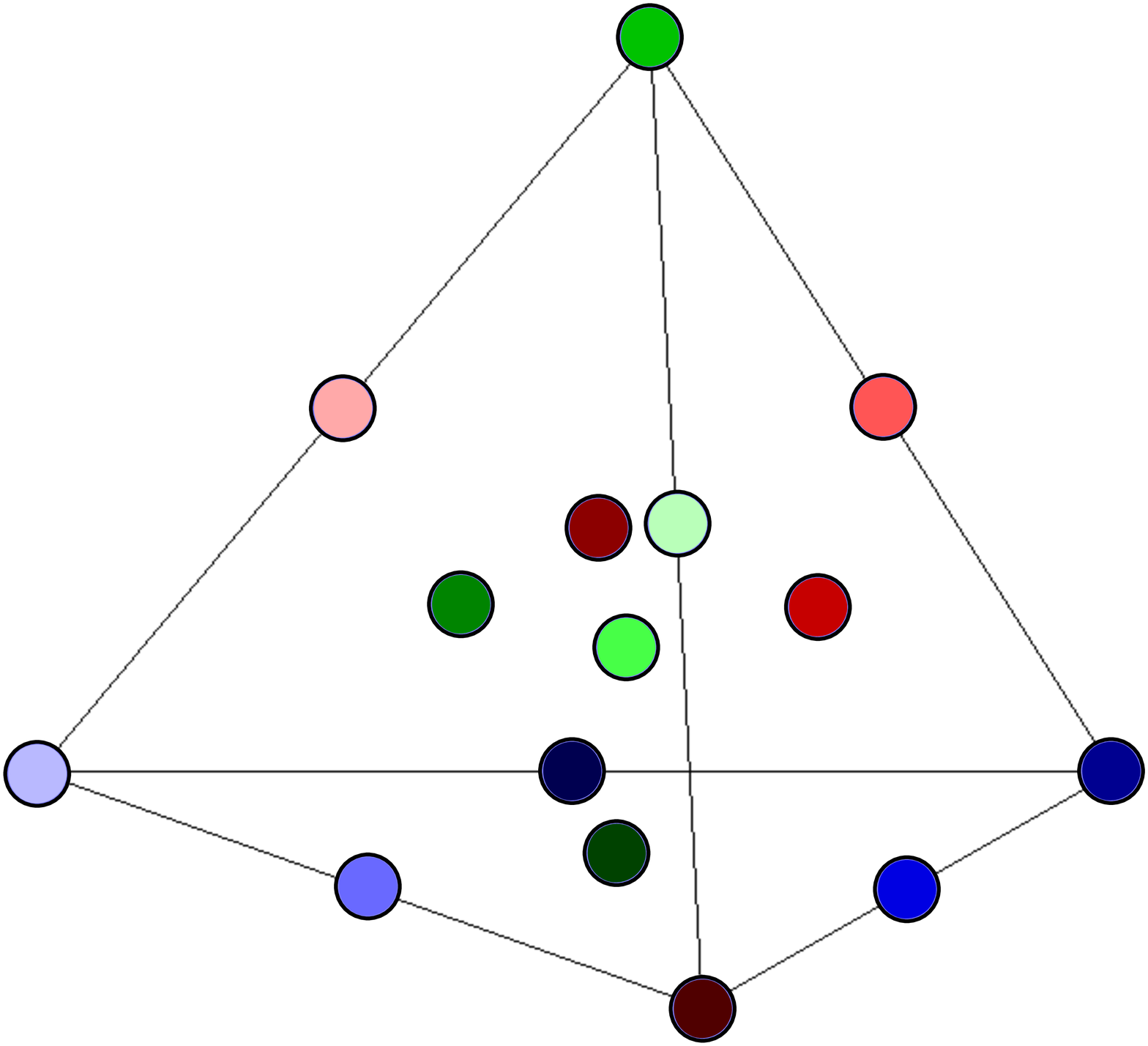}}
\caption{A Kirkman color design. The elements in this KTS represent the solution to the design $D(15,3,1)|Q_{12},Q_{10},Q_4,Q_{11}\rangle$.}
\end{figure}
%\end{minipage}
% ==============================================================================================================

\begin{table*}
\begin{center}
\begin{tabular}{||c|c|c||c|c|c||c|c|c||c|c|c||c|c|c||c|c|c||c|c|c||}

\hline
 &SUN& & &MON& & &TU& & &WED& & &TH& & &FRI& & &SAT& \\
 \hline
 &111& & &001& & & 010& & &011& & &100& & &101& & & 110& \\
 \hline  \hline
$R_0$&$G_0$&$B_0$&$B_2$&$R_0$&$B_1$&$G_0$&$B_1$&$B_3$&$B_3$&$B_2$&$B_0$&$G_0$&$B_2$&$B_4$&$B_4$&$B_3$&$R_0$&$B_1$&$B_0$&$B_4$\\
 \hline $R_1$&$B'_2$&$G_1$&$R_2$&$B'_0$&$R_1$&$R_2$&$R''_0$&$R_3$&$R_4$&$R'_0$&$R_1$&$R_3$&$B_0$&$R_4$&$R_1$&$G_0$&$R_3$&$R_4$&$G_0$&$R_2$\\
 \hline $R_3$&$B_1$&$G_2$&$G_1$&$G_0$&$G_4$&$G_1$&$B_0$&$G_2$&$G_3$&$G_0$&$G_2$&$G_2$&$R_0$&$G_4$&$G_4$&$B'_0$&$G_3$&$G_3$&$R'''_0$&$G_1$\\
 \hline $R_2$&$B'''_4$&$G_3$&$G_2$&$B_4$&$R_4$&$R_4$&$B'_2$&$G_3$&$G_1$&$B'''_4$&$R_3$&$G_1$&$B'_3$&$R_2$&$G_1$&$B'_1$&$R_4$&$G_2$&$B_3$&$R_1$\\
 \hline $R_4$&$B_3$&$G_4$&$G_3$&$B'_3$&$R_3$&$G_4$&$B'_4$&$R_1$&$G_4$&$B''_1$&$R_2$&$G_3$&$B'''_1$&$R_1$&$G_2$&$B_2$&$R_2$&$G_4$&$B_2$&$R_3$\\
\hline
\end{tabular}
\end{center}
\caption{ A Kirkman arrangement of five rows of three for each of seven days of the week. The binary label for each day also specifies that of the central entry of the first row for that day. 15 of the triplets are commuting and 20 cyclic and for the latter have been arranged in cyclic order. All rows of SUN and the lowest two rows of the other days are colorless combinations of BGR. Considering flavor zero as able to take on the color of the two other partners, the first three rows of MON through SAT are blue, red, and green, respectively. This design comes from our running example $D(15,3,1)|Q_{12},Q_{10},Q_4,Q_{11}\rangle$.}
\end{table*}

The power of the color scheme is evident from Table IV. All five rows of Sunday and the bottom two rows of all days of the week are colorless combinations of one each of [B, R, G]. The top three rows of the six days besides Sunday are pure colors B, G, and R, respectively, all three sharing the same color. For this purpose, we will interpret for flavor 0 a chameleon-like ability to take on the common color of the other two in the triplet! Finally, Fig. 6 re-renders the tetrahedron of Fig. 3 with all fifteen points in the same color-flavor scheme. Our color-flavor scheme bears some resemblance to the ``color hats guessing game" of \cite{ref12} which, however, used only two different colored hats. Another connection noted there is to the Hamming code, KTS as a (15, 11) Hamming code. Yet another color scheme that can be implemented algorithmically to generate designs will be presented in the sub-section below.

\begin{figure}
\scalebox{2.0}{\includegraphics[width=3in]{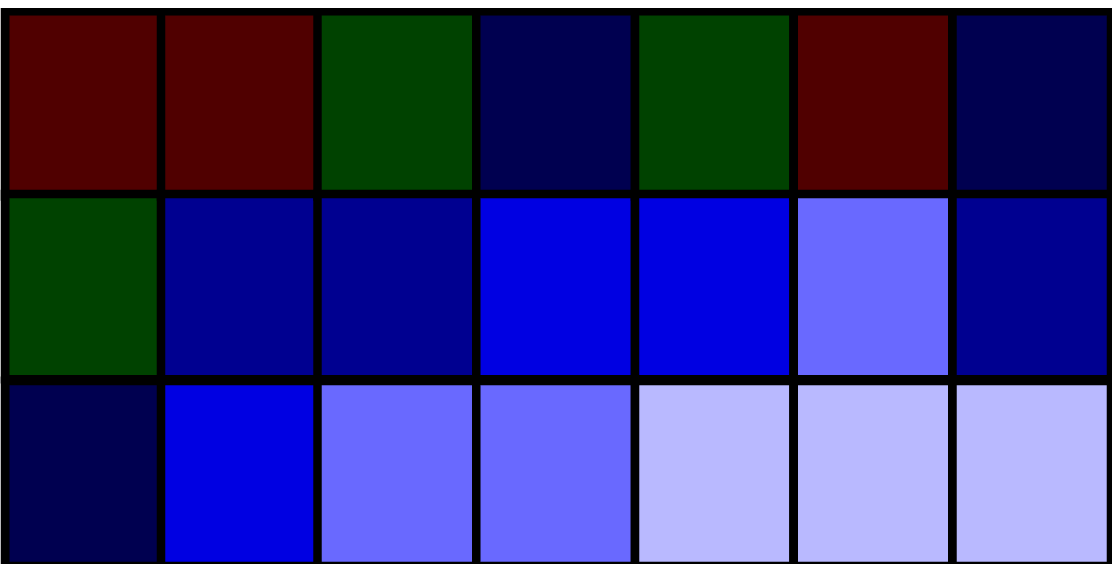}}
\caption{$D_c(7,3,1)|Q_{10},Q_4,Q_{11}\rangle$}
\end{figure}
  
%\fig[width=7in]{Images/Fig8.eps}
\begin{figure}
\scalebox{2.0}{\includegraphics[width=3in]{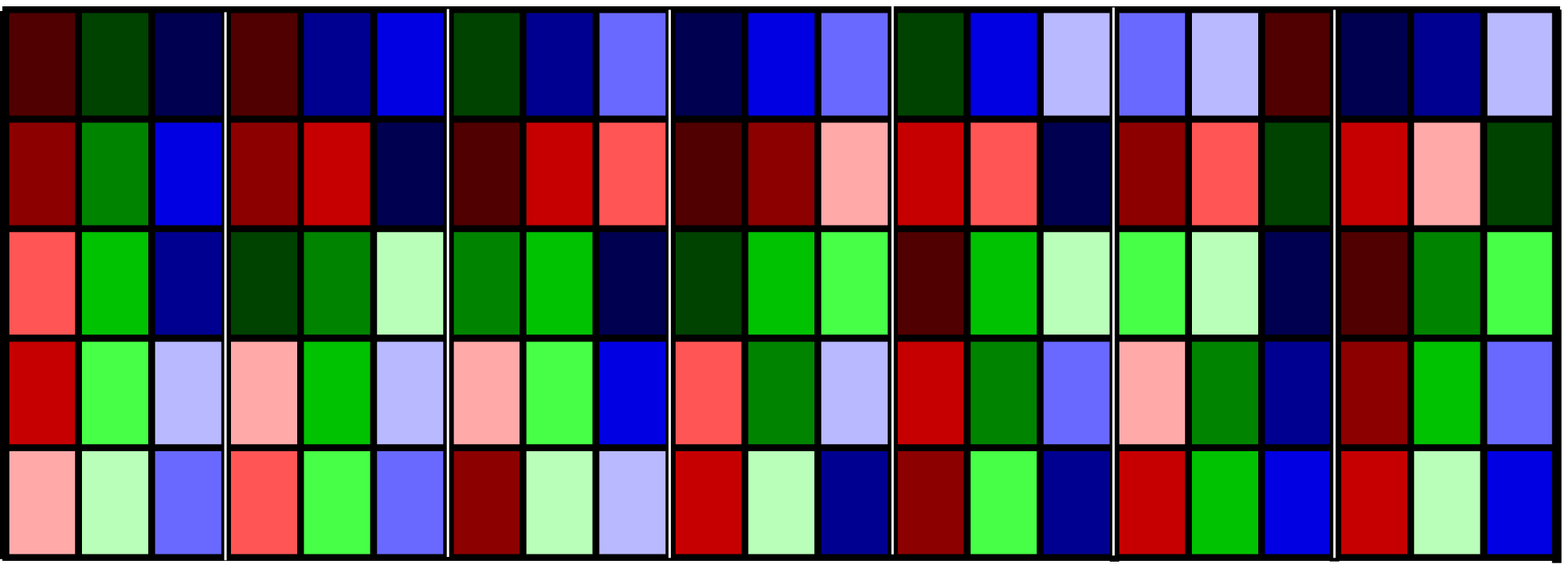}}
\caption{$K_c(15,3,1)|Q_{12},Q_{10},Q_4,Q_{11}\rangle$} 
\end{figure}

As already noted, color has played an enormous role in human perception, art, and philosophy in many cultures. The very naming of a color poses deep questions that have been addressed in literature \cite{ref31}. Using the color-flavor language hereto developed, we can cast resolved design systems into color tilings. To construct a $(v,k,\lambda)$ tiling, we assign a subset of the $v$ colors of the design to the $k$ tiles in each block, and the $v$ colors can be read directly from a resolved operator design with the use of our dictionary in Table I. Hence, in Fig. 7, we see a color design that is a representation of the $D(7,3,1)|Q_{10},Q_{4},Q_{11}\rangle$ example used throughout this paper. Furthermore, if a Kirkman design $(3n,3,1)$ is of odd $n$, we can find a resolved design wherein the $v$ varieties of the design appear once in a number of discrete sets or ``days''. For such a resolved Kirkman system, we construct rows of colored tiles packed into sets of all $v$ colors. Hence, in Fig. 8, we see a Kirkman design that is a representation of the $D(15,3,1)|Q_{12},Q_{10},Q_{4},Q_{11}\rangle$ example used throughout this paper.

\section{Acoustic Designs}

A single note, tone, or other basic acoustic element can equally serve as the primitive objects on which a design operation such as the ones we have set up acts. The sensation of sound, its role in speech, hearing and music, makes it equally as important in human life and civilization as the sensation of light and color discussed in previous sections. A major difference lies in the number of tonal bases, which is even larger than the 3 color-bases the human eye can sense. This richer number of bases is seen in the larger basis-set for scales like the chromatic scale or the 15-note scale. For acoustic oscillations, harmonic analysis ever since Fourier has used decomposition into single frequencies as forming a basis that is conjugate to a description in moments of time. Western musical notation, on the other hand, has long employed a ``mixed system" featuring both conjugate quantities of frequency and time, the former in the five bars on which notes are depicted while time is laid out horizontally. Gabor \cite{ref32} proposed an alternative to Fourier's analysis in terms of ``wavelets" that may be regarded as ``acoustic quanta," with partial localization in both time and frequency, and these are more powerfully adapted to applications in music and science.  

In place of the color-bases of the previous section, we now form the basic elements of the design from these acoustic quanta, and instead of producing designs of color, we now find acoustic signals that represent solutions of the design system through operations discussed in \cite{ref25}. Immediately, there is much more freedom in choosing the tonal-bases than was found with color-bases. To form our quanta, we construct wavelets that are the product of a windowing function centered at different moments in time and a harmonic oscillation centered at different frequency bands. This partial localization in time and frequency allows a mapping to a lattice of points on the time-frequency plane, with the design elements occupying localized areas that are bound by uncertainty principles. Each design element takes on a unique time-frequency coordinate. Elements in a triplet share the same temporal coordinate such that they sound together in a chord when a signal is played. Hence, it is the temporal occurrence of a note that gives information about the location of the design element in the block design. Conjugate to the temporal aspect, the harmonic aspect of a design element describes the identity of the operator within the triplet. With each of the fifteen notes in the scale as one of our fifteen operators, the frequencies of our design elements span a 15-note scale such that the element with lowest index $i$ in $O_i$ notation rings at the lowest frequency of the scale and each subsequent element rings at increasingly higher frequencies in the scale. 

At this point, a notation for the elements of an acoustic design becomes useful. Western musical notation assigns the seven letters $[A, B, ..., G]$ with accompanying signature as sharp ($\sharp$), natural (no symbol), or flat ($\flat$) to label the pitches of the chromatic scale. The number of possible combinations with such a scheme is, of course, larger than 12, and some symbol combinations make reference to the same musical pitch. For example, at the formal level of notation, $C^{\sharp}$ is the same as $D^{\flat}$ and $E^{\sharp}$ is the same as $F$. Music theorists call this degeneracy in note names and pitch ``enharmonic equivalence". To label the elements of our fifteen note scale, we construct a similar notation using a set of five letters $[A, B, C, D, E]$ along with the three signatures of sharp, neutral, or flat. However, in our scheme each combination of symbols represents a unique pitch such that there is no enharmonic equivalence. This is, in some sense, an inversion of the color-flavor labeling constructed for design of color, with the three types of color now acting like the three types of signatures and five types of flavors acting like the five letters. We append one more row to the dictionary in Table I to include these musical labels.  

Having now a mapping from basic operators to tonal-bases, we can transcribe our previous design of operators and colors into sounds. Using our dictionary in Table I, we can read seven chords from seven lines in the Fano design of Fig. 4. Each chord contains three notes that correspond to the three design elements that sit on each line, and the identity of the note is given in our mapping from color-base to tonal-base. In this way, the center altitude of Fig. 4 is read as a chord of the notes $B$, $D$, and $E$. For this simple KTS, the order of the chords does not matter. The design in Fig. 4 produces a signal that plays seven chords at seven moments of time, such that each chord contains three notes and no two notes appear together more than once in any chord. 

Thus, we also read a signal from the Kirkman design in Fig. 6, which renders a design signal of 35 chords. Each chord again contains three notes such that no pair of notes appears together more than once in a chord. The notes appear with a repetition number of seven so that we can resolve the triplets of this KTS system into an arrangement of five rows across seven days, finding a solution to Kirkman's design. This would replace in Table IV the elements in the last row by a one-to-one matching of the last two rows of Table I, thereby producing a sequence of chords that begin with the first triplet of Sunday and proceed down the rows of the design until the last triplet of Saturday sounds. 

%\begin{table*}
%\begin{center}
%\begin{tabular}{||c|c|c||c|c|c||c|c|c||c|c|c||c|c|c||c|c|c||c|c|c||}

%\hline
 %&SUN& & &MON& & &TU& & &WED& & &TH& & &FRI& & &SAT& \\
 %\hline
 %&111& & &001& & & 010& & &011& & &100& & &101& & & 110& \\
 %\hline  \hline
%$D$&$B$&$E$&$E^{\sharp}$&$D$&$B^{\flat}$&$B$&$B^{\flat}$&$A^{\flat}$&$A^{\flat}$&$E^{\sharp}$&$E$&$B$&$E^{\sharp}$&$D^{\flat}$&$D^{\flat}$&$A^{\flat}$&$D$&$B^{\flat}$&$E$&$D^{\flat}$\\
 %\hline %$E^{\flat}$&$E^{\sharp}$&$A$&$A^{\sharp}$&$E$&$E^{\flat}$&$A^{\sharp}$&$D$&$C$&$B^{\sharp}$&$D$&$E^{\flat}$&$C$&$E$&$B^{\sharp}$&$E^{\flat}$&$B$&$C$&$B^{\sharp}$&$B$&$A^{\sharp}$\\
 %\hline %$C$&$B^{\flat}$&$D^{\sharp}$&$A$&$B$&$C^{\flat}$&$A$&$E$&$D^{\sharp}$&$C^{\sharp}$&$B$&$D^{\sharp}$&$D^{\sharp}$&$D$&$C^{\flat}$&$C^{\flat}$&$E$&$C^{\sharp}$&$C^{\sharp}$&$D$&$A$\\
 %\hline %$A^{\sharp}$&$D^{\flat}$&$C^{\sharp}$&$D^{\sharp}$&$D^{\flat}$&$B^{\sharp}$&$B^{\sharp}$&$E^{\sharp}$&$C^{\sharp}$&$A$&$D^{\flat}$&$C$&$A$&$A^{\flat}$&$A^{\sharp}$&$A$&$B^{\flat}$&$B^{\sharp}$&$D^{\sharp}$&$A^{\flat}$&$E^{\flat}$\\
 %\hline %$B^{\sharp}$&$A^{\flat}$&$C^{\flat}$&$C^{\sharp}$&$A^{\flat}$&$C$&$C^{\flat}$&$D^{\flat}$&$E^{\flat}$&$C^{\flat}$&$B^{\flat}$&$A^{\sharp}$&$C^{\sharp}$&$B^{\flat}$&$E^{\flat}$&$D^{\sharp}$&$E^{\sharp}$&$A^{\sharp}$&$C^{\flat}$&$E^{\sharp}$&$C$\\
%\hline
%\end{tabular}
%\end{center}
%\caption{ A Kirkman arrangement of five rows of three for each of seven days of the week identical to the $D(15,3,1)|%Q_{12},Q_{10},Q_4,Q_{11}\rangle$ design found in Table IV, now cast in our musical notation.}
%\end{table*}

\subsection{Constructing acoustic designs}

The mathematical nature of music has long been of interest to philosophers. Modern literature on musical theory addresses the geometric and algebraic nature of the complicated harmonies in music \cite{ref33}. We consider only a narrow view for the purpose of rendering acoustic designs in this study. Following Gabor's construction of acoustic quanta as products of windowing functions and harmonic oscillations, we choose our wavelets, and we take special care to construct wavelets that when sounded together produce distinct chords when heard by the human ear. 

We first address the windowing function, which can be written $W_n(t)=W(t-nh)$, where $n$ denotes the temporal displacement and $h$ denotes the hop size of the windowing function. There are a number of possible choices for the windowing function, including the square window, the B-spline, the Cosine window, and Gaussian function. Each type of windowing function introduces artifacts in the frequency spectrum of a note, yet we find that, as long as the windowing function is sufficiently smooth, the structure of the window has only a small effect on the way that notes sound together in chords, this being rather a consequence of the harmonic oscillations modulating each window. Thus, we choose the Gaussian function to window our notes and we adjust the width of the Gaussian so that the onsets and decays of the chords can be distinguished. 

Because the harmonic aspect of the wavelet has the most significant effect on the distinguishability of different tonal-bases, we must be careful with the definition of our 15-note scale. Inspired by the chromatic scale, we might be tempted to choose 15-note equal temperament. However, we will find that equal temperament will not provide a set of notes that are readily distinguishable by the human ear. The problem lies in the fact that the notes of an equal tempered scale are not very consonant among themselves, and so they will produce chords that all sound very similar. It is for this reason that western music generally works with a sub-set of the chromatic scale, such as the C-major scale, to create music. In order to maximize the distinguishability of the notes, we must turn to some non-traditional ideas in music theory. One such idea is the use of ``combination product sets" (CPSs) for scale production, which Ervin Wilson used to construct his famous ``Hexany" system \cite{ref34}. To construct a CPS, one first picks a number $N$ of prime numbers, excluding two as the octave number, and a combinatoric number $M$. The octave divisions of the scale are then found by multiplying $M$ prime numbers and then rescaling the product by a power of two such that the elements lie on a range $[1,2]$ in the octave. A `tonic' is then chosen for the scale, which defines the frequency of the lowest note. All further frequencies are then given by the product of the tonic and the other scaled products. We employ this method to construct our scale, using a CPS$(6,2)$ for the 15-note scale. 

Using 15 tonal bases as a 15-note scale, the Fano (7, 3, 1) design rendered previously in color in Fig. 4 and Fig. 7 takes the acoustical form shown in Fig. 9 and the Kirkman (15, 3, 1) design of Figs. 6 and 8 the form shown in Fig. 10. Note how the (7, 3, 1) design in Fig. 9 is a subset formed by the first vertical bars of each of the seven day panels in the (15, 3, 1) design of Fig. 10, an aspect of the nesting of the Fano triangle's 2-simplex of Fig. 4 within the tetrahedron's 3-simplex of Fig. 6 discussed earlier. 

\begin{figure}
\scalebox{2.0}{\includegraphics[width=1in]{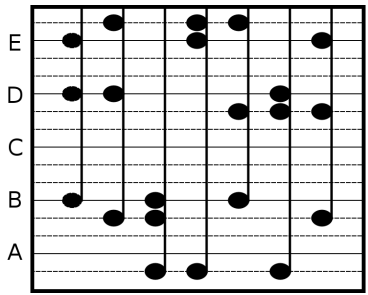}}
\caption{$D_s(7,3,1)|Q_{10},Q_4,Q_{11}\rangle$} 
\end{figure}
%\fig[width=5in]{Images/Fig9.eps}
%\end{center} 
\begin{figure}
\scalebox{2.0}{\includegraphics[width=3in]{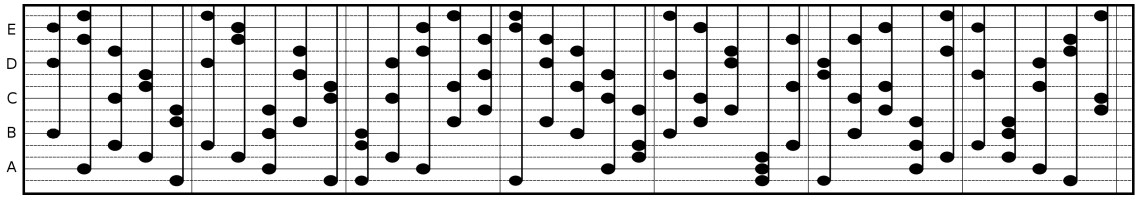}}
\caption{$K_s(15,3,1)|Q_{12},Q_{10},Q_4,Q_{11}\rangle$}
%\fig[width=7in]{Images/Fig10.eps} 
\end{figure}

\section{Summary and Conclusion}

In a remarkable fashion, a mathematical puzzle has had a recurring and long standing impact on mathematics and physics for nearly two centuries, and further connections seem to continue to appear. This paper has discussed one such to the states and operators of two quantum spins or qubits that lie at the core of the field of quantum information. Using their multiplication properties that physicists are very familiar with, we have presented a systematic method for generating solutions to the problem posed by the puzzle and translated them to designs in color and acoustic patterns.

 We end by noting yet further connections and possibly useful avenues for linking disparate topics because the SU(4) symmetry group of 15 elements considered here has close similarities to the SO(4, 2) group that is the full symmetry group of the quantum-mechanical description of the hydrogen atom and is also the Poincare group of space-time symmetries \cite{ref35}.

\end{document}